\documentclass[10pt, a4paper, onecolumn]{article}
\usepackage{ifthen}
\usepackage[T1]{fontenc}
\usepackage[ansinew]{inputenc}
\usepackage[francais]{babel}
\usepackage{amssymb}
\usepackage{eucal}
\usepackage{epsf}
\DeclareMathAlphabet\Dbb{U}{bbold}{m}{n}
\begin{document}

\title{Magneto-electro-optical properties of the quantum vacuum and Lorentz
invariance}
\author{C. Rizzo$^{1}$ and G.L.J.A. Rikken$^{2}$ \\
\\
$^{1}$Laboratoire Collisions Agr\'{e}gats R\'{e}activit\'{e}, \\
Universit\'{e} Paul Sabatier/CNRS, \ F-31062 Toulouse, France\\
$^{2}$Laboratoire National des Champs Magn\'{e}tiques Puls\'{e}s, \\
CNRS/INSA/UPS, BP4245, F-31432 Toulouse, France.}
\maketitle

\begin{abstract}
We consider the magneto-electric optical properties of the quantum
vacuum and show that all the different phenomenona are related by
Lorentz invariance. As a model calculation we show how crossed
fields properties can be calculated starting from single field
properties by using Lorentz transformations. Using this method we
have studied for the first time the case of a crossed static
magnetic field and electric field applied with one of these two
fields parallel to the direction of light propagation. We also
show that parallel field properties can be found using general
symmetry properties.

PACS: 42.25.Lc, 12.20.Ds \medskip \newline \textit{Corresponding
author}: C. Rizzo LCAR \newline
\end{abstract}

\subsection*{Introduction}

Several new magneto-electric optical phenomena in centrosymmetric
media have
recently been observed for the first time \cite{RandR1} \cite{RandR2} \cite%
{RSW-PRL}. When a static magnetic field $\mathbf{B}_{0}$ and electric field $%
\mathbf{E}_{0}$ are applied perpendicular to each other and to the
propagation vector of the light $\mathbf{k}$, the existence of
magneto-electric linear birefringence \cite{RandR2}, and the
existence of a polarization-independent anisotropy, proportional
to $\mathbf{B}_{0}\times
\mathbf{E}_{0}$ \cite{RSW-PRL} have been proven. When, on the other hand, $%
\mathbf{B}_{0}$ and $\mathbf{E}_{0}$ are parallel, the existence
of magneto-electric Jones birefringence was demonstrated for the
first time \cite{RandR1}.

The most elementary centrosymmetric medium is the quantum vacuum.
Non-linear optical phenomena in vacuum have been predicted since
1935 \cite{Euler} in the framework of quantum electrodynamics
\cite{Jackson} \cite{Milloni}. In particular, the existence in the
vacuum of a linear birefringence induced by a transverse static
magnetic field (the Cotton-Mouton effect), or by a
transverse static electric field (the Kerr effect), has been predicted \cite%
{Bialynicka}, but not yet observed. Recently, magneto-electric
linear birefringence, magneto-electric Jones birefringence
\cite{RikkenandRizzo}
and polarization-independent magneto-electric anisotropy \cite%
{RikkenandRizzoBR} have also been predicted for the quantum
vacuum.

Here we perform a complete study of the magneto-electro-optical
properties of the quantum vacuum using Lorentz invariance. Using a
suggestion by the authors of Ref. \cite{RSW-PRL}, we relate single
field properties to crossed field properties like magneto-electric
anisotropy. This method is based on the magnetic (electric) field
behavior under Lorentz transformation that allows us to look for
an appropriate reference frame in which the two crossed fields are
transformed into only one field. In this reference frame the only
existing effect is a magnetic (electric) linear birefringence
whose characteristic values of index of refraction are known.
These refractive index values can be transformed back to the
laboratory frame values by using the well-known result of the
electrodynamics of moving media, as e.g. used to describe Fizeau's
experiment \cite{Jackson}. This method is ideally suited for the
quantum vacuum since it is invariant under Lorentz transformation,
and one can calculate the analytical value of the refractive index
in any reference frame.

\subsection*{Magneto-electro-optics of the quantum vacuum}

With this method, we reproduce in a straightforward way the
results, obtained in previous publications by more elaborate
calculations, concerning magneto-electric birefringences, and
polarization-independent anisotropy, proportional to
$\mathbf{B}_{0}\times \mathbf{E}_{0}$. The case where a
static magnetic field $\mathbf{B}_{0}$ and a static electric field $\mathbf{E%
}_{0}$ are simultaneously applied perpendicular to each other but
with one of these two static fields parallel to the direction of
propagation has never been treated before. Here we demonstrate for
the first time that only the field perpendicular to the direction
of propagation gives an effect and that no bilinear phenomenon
exists for this geometry. In the same context we also examine the
case of both fields parallel to the direction of propagation. By
symmetry arguments we relate parallel fields properties to crossed
fields properties and show that no effect exists for this
geometry.

The starting point of any calculation of the propagation of light
in the quantum vacuum \cite{RikkenandRizzo},
\cite{RikkenandRizzoBR} is the Heisenberg-Euler Lagrangian
\cite{HeisenbergEuler}. The form of the effective Lagrangian
$L_{HE}$ of the electromagnetic interaction is essentially
determined by the fact that the Lagrangian has to be relativistic
and CPT invariant and therefore can only be a function of the
Lorentz invariants \cite{Landau} F, G that in Heaviside-Lorentz
units can be written as
\begin{equation}
F=(E^{2}-B^{2})
\end{equation}%
\begin{equation}
G=(\mathbf{{E}\cdot {B})}
\end{equation}%
Up to second order in the fields, $L_{HE}$ can be written as $%
L_{HE}=L_{0}+L_{EK}$ where $L_{0}$ is the usual Maxwell`s term ${\frac{1}{2}}%
F$ and $L_{EK}$ is the first order non linear term first
calculated by Euler and Kockel \cite{Euler}. $L_{EK}$ is valid in
the approximation that the fields vary slowly over the Compton
wavelength of the electron $\lambda
=\hbar /m_{e}c$ and during a time ${{t_{e}}=}\lambda /c$ . Moreover $E$ and $%
B$ have to be smaller than the critical field
$E_{cr}=m_{e}^{2}c^{3}/e\hbar $ i.e. $B\ll 4.4\times 10^{9}$ T and
$E\ll 1.3\times 10^{18}$ V/m. $L_{HE}$ can be written in
Heaviside-Lorentz units as
\begin{equation}
L_{HE}=L_{0}+{\frac{1}{2}}(aF^{2}+bG^{2})
\end{equation}%
A term proportional to $FG$ is Lorentz invariant but not CPT
invariant and therefore does not appear in the expression of
$L_{EK}$. Higher order terms of $L_{HE}$ can be written in the
same way by looking for combinations of the two invariants $F$ and
$G$ that also respect CPT. In the case of a plane wave in vacuum,
both $F$ and $G$ are equal to zero. The propagation of a plane
wave
in vacuum is thereby not affected by non linear interactions since $L_{HE}=0$%
.

The values of $a$ and $b$ are provided by QED. The calculation by
Heisenberg
and Euler gives $a=e^{4}\hbar /45\pi m^{4}c^{7}=2.67\ 10^{-32}\ G^{-2}$ and $%
b=7a$. Based on this Lagrangian, in the case of crossed static fields $%
\mathbf{E}_{0}$ and $\mathbf{B}_{0}$, we showed in Ref.
\cite{RikkenandRizzo} the existence of a Cotton-Mouton
birefringence $\Delta n_{CM}\propto B_{0}^{2}$, a Kerr
birefringence $\Delta n_{K}\propto E_{0}^{2}$ and a
magneto-electric linear birefringence $\Delta n_{MELB}\propto
E_{0}B_{0}$. In Ref. \cite{RikkenandRizzoBR} we also showed the
existence of a magneto-electric anisotropy, independent of
polarization, $\Delta n_{MEA}$ which is also proportional to
$E_{0}B_{0}$. In the case of parallel static fields we also showed
in Ref. \cite{RikkenandRizzo} the existence of a magneto-electric
Jones birefringence corresponding to a $\Delta n_{J}\propto
E_{0}B_{0}$. Faraday and Pockel effects are not permitted in
vacuum since no terms containing three electromagnetic fields
exist in the Heisenberg-Euler Lagrangian. As far as we know, the
case of crossed electric and magnetic fields with one of the two
parallel to the direction of propagation of light has never been
treated, neither in vacuum, nor in any other medium. All the
effects predicted for the quantum vacuum have been observed in
centrosymmetric media. The quantum vacuum behaves exactly like any
other centrosymmetric medium. The predicted values for all these
effects in vacuum are unfortunately so small that observation has
not yet been possible.

For our purpose, we just need to suppose that when a static magnetic field $%
\mathbf{B}_{0}$ is present in a vacuum, perpendicular to the
direction of light propagation, the light velocity changes in such
a way that
\begin{equation}
n_{\Vert }=1+\eta_{\Vert }{B}_{0}^{2}  \label{npar}
\end{equation}%
and%
\begin{equation}
n_{\bot }=1+\eta_{\bot}{B}_{0}^{2}  \label{nperp}
\end{equation}%
where $n_{\Vert }$ and $n_{\bot }$ are the index of refraction for
light polarized parallel and orthogonal to the static magnetic
field,
respectively. Since the velocity of light has to be smaller than $c$, $%
\eta_{\Vert }$ and $\eta_{\bot}$ are positive.

\subsection*{E and B fields under Lorentz transformations}

The case of two crossed fields $\mathbf{E}_{0}$ and
$\mathbf{B}_{0}$ can be completely solved by an appropriate
Lorentz transformation thanks to the Lorentz invariance of the
quantum vacuum. Using Lorentz transformations one
can express the fields $\mathbf{E}$ and $\mathbf{B}$ in an inertial frame $%
K^{^{\prime }}$ in terms of the values in another inertial frame
$K$. For a general Lorentz transformation from frame $K$ to a
frame $K^{^{\prime }}$ moving with velocity {\boldmath
$\beta$}$=\mathbf{v/}c$ relative to $K$, the
transformation of the fields can be written \cite{Becker}%
\begin{equation}
\mathbf{E}^{\prime }=\gamma (\mathbf{E}+\mbox{\boldmath $\beta$ }\times \mathbf{B})-{%
\frac{\gamma ^{2}}{{\gamma +1}}}\mathbf{\beta }(\mbox{\boldmath $\beta$ }\cdot \mathbf{%
E})  \label{LorTransE'}
\end{equation}

\begin{equation}
\mathbf{B}^{\prime }=\gamma (\mathbf{B}-\mbox{\boldmath $\beta$ }\times \mathbf{E})-{%
\frac{\gamma ^{2}}{{\gamma +1}}}\mbox{\boldmath $\beta$ }(\mbox{\boldmath $\beta$ }\cdot \mathbf{%
B})  \label{LorTransB'}
\end{equation}%
where $\mathbf{E}$ and $\mathbf{B}$ are the electric and magnetic
fields in the frame $K$, $\mathbf{E}^{\prime }$ and
$\mathbf{B}^{\prime }$ are the electric and magnetic fields in the
frame $K^{^{\prime }}$ and $\gamma \equiv \left( 1-\beta
^{2}\right) ^{-1/2}$ as usual. Let's suppose that $\mathbf{E}$ and
$\mathbf{B}$ are perpendicular with $E>B$. It is possible to
choose the velocity $\mathbf{v}$ so that in the corresponding
frame $K^{^{\prime }}$
only the electric field $\mathbf{E}^{\prime }$ exists. To calculate $\mathbf{%
v}$ it is sufficient to put $B^{^{\prime }}=0$ in Eq.
$\ref{LorTransB'}$ and to remark that a solution can be found only
if $\mathbf{v}$ is perpendicular
to $\mathbf{E}$ and $\mathbf{B}$. The result is that%
\begin{equation}
\mathbf{v}=c{\frac{\mathbf{E}\times \mathbf{B}}{E^{2}}}
\end{equation}%
and therefore $\beta =B/E$. It is evident that this solution is
only valid if $E>B$. In the case when $B>E$, one can find a
velocity $\mathbf{v}$ so
that in the corresponding frame $K^{^{\prime }}$ only the magnetic field $%
\mathbf{B}^{\prime }$ exists. In this case%
\begin{equation}
\mathbf{v}=c{\frac{\mathbf{E}\times \mathbf{B}}{B^{2}}}
\end{equation}%
and therefore $\beta =E/B$. The value of $\mathbf{E}^{\prime }$
for $E>B$ (resp. of $\mathbf{B}^{\prime }$ for $B>E$) can be
calculated by inserting the
corresponding value of $\mathbf{v}$ in Eq. \ref{LorTransE'} (resp. Eq. \ref{LorTransB'}).
One obtains%
\begin{equation}
\mathbf{E}^{\prime }=\sqrt{1-\beta ^{2}}\mathbf{E}
\end{equation}%
and%
\begin{equation}
\mathbf{B}^{\prime }=\sqrt{1-\beta ^{2}}\mathbf{B}
\end{equation}%
respectively.

\subsection*{Light propagation in a moving medium}

The basic formulas to study the propagation of light in a moving
frame are the ones that give the magnitude and the direction of a
velocity $\mathbf{u}$ obtained by adding relativistically two
velocities $\mathbf{u}^{\prime }$
and $\mathbf{v}$ \cite{Becker}. In particular, one finds that%
\begin{equation}
u^{2}={\frac{u^{^{\prime }2}+v^{2}+2u^{^{\prime }}v\cos(\theta
^{^{\prime
}})-({\frac{u^{^{\prime }}v}{c}})^{2}\sin(\theta ^{^{\prime }})^{2}}{(1+{%
\frac{u^{^{\prime }}v}{c^{2}}}\cos(\theta ^{^{\prime }}))^{2}}}
\label{GenTransform}
\end{equation}%
and%
\begin{equation}
\tan(\theta )={\frac{u^{^{\prime }}\sin(\theta ^{^{\prime }})\sqrt{1-\beta ^{2}%
}}{{u^{^{\prime }}\cos(\theta ^{^{\prime }})+v}}}
\label{TransformAngles}
\end{equation}%
where $\theta ^{^{\prime }}$ is the angle between
$\mathbf{u}^{^{\prime }}$ and $\mathbf{v}$ and $\theta $ the angle
between $\mathbf{u}$ and $\mathbf{v} $. To apply these formulas to
our calculation, let's write $u^{^{\prime }}=c/n^{\prime }$ and
$u=c/n$, $v$\textbf{\ }being the moving frame velocity. In
particular, if $\theta ^{^{\prime }}=0$, $\theta =0$ and moreover
if $\beta <<1$, one obtains the well-known Fizeau formula

\begin{equation}
u={\frac{c}{n^{^{\prime }}}}+v(1-{\frac{1}{n^{^{\prime }2}}})
\end{equation}%
For the quantum vacuum $n$ and $n^{^{\prime }}$ can be written as $%
n=1+\delta n$ with $\delta n<<1$ and $n^{^{\prime }}=1+\delta
n^{^{\prime }}$ with $\delta n^{^{\prime }}<<1$. If $\mathbf{u}$
and $\mathbf{v}$ are parallel, $\theta =0$ and $\theta ^{^{\prime
}}=0$ so that Eq. $\ref{GenTransform}$ can be written, up to first
order with respect to $\delta n$ and $\delta n^{^{\prime
}} $, as%
\begin{equation}
\delta n=\delta n^{^{\prime }}{\frac{{1-\beta }}{{1+\beta }}}
\label{Fizeau1}
\end{equation}%
If $\mathbf{u}$ and $\mathbf{v}$ are antiparallel, one has simply
to change
the sign of $\beta $. If $\mathbf{u}$\textbf{\ }is perpendicular to $\mathbf{%
v}$, $\theta ={\frac{\pi }{2}}$. From Eq. $\ref{TransformAngles}$,
we can infer that $\cos(\theta ^{^{\prime }})=-{v/u}^{\prime }$
and $\sin(\theta ^{^{\prime }})=\sqrt{1-({v/u}^{\prime })^{2}}$.
This obviously means that in $K^{^{\prime }}$, the direction of
propagation of light is no longer
perpendicular to the frame velocity. Upon inserting the values of $%
\cos(\theta ^{^{\prime }})$ and $\sin(\theta ^{^{\prime }})$ into Eq. $\ref%
{GenTransform}$, one obtains%
\begin{equation}
\delta n=\delta n^{^{\prime }}{\frac{1}{{1-\beta ^{2}}}}
\label{Fizeau3}
\end{equation}%

\subsection*{Light polarization in a moving medium}

We are interested in effects that depend on the polarization of
light. We therefore have to study how the polarization of light is
transformed going from frame $K$ to frame $K^{^{\prime }}$. We
first consider the case of a plane wave in $K$ when no external
fields are present and recall that the quantities $E^{2}-B^{2}$
and $\mathbf{E}\cdot \mathbf{B}$ are invariant under Lorentz
transformation. For a plane wave in vacuum $E_{\omega
}=B_{\omega }$ and $\mathbf{E}_{\omega }\cdot \mathbf{B}_{\omega }=0$ where $%
\mathbf{E}_{\omega },\mathbf{B}_{\omega }$ are the optical fields. Since $%
E_{\omega }^{2}-B_{\omega }^{2}$ and $\mathbf{E}_{\omega }\cdot \mathbf{B}%
_{\omega }$ are equal to zero in $K$, they are also equal to zero in $%
K^{^{\prime }}$. This means that in $K^{^{\prime }}$ we still have
a plane wave. If in addition, static $\mathbf{E}_{0}$ and
$\mathbf{B}_{0}$ fields
are also present, the total fields can be written as $\mathbf{E}=\mathbf{E}%
_{0}+\mathbf{E}_{\omega }$ and
$\mathbf{B}=\mathbf{B}_{0}+\mathbf{B}_{\omega }$. Using the
linearity of Eqs.
\ref{LorTransE'} and \ref{LorTransB'} with respect to the fields, it is straightforward to show that%
\begin{equation}
\mathbf{{E}^{\prime }={E}_{0}^{\prime }+{E}_{\omega }^{\prime }}
\end{equation}

\begin{equation}
\mathbf{{B}^{\prime }={B}_{0}^{\prime }+{B}_{\omega }^{\prime }}
\end{equation}%
where $\mathbf{E}_{\omega }^{\prime }$\ and $\mathbf{B}_{\omega
}^{\prime }$ are the transformations of the optical fields alone.
We can therefore conclude that the optical fields transform as if
no external field were present and thus a plane wave in $K$
remains a plane wave in $K^{^{\prime }}$. Let's now investigate
how the orientation of $\mathbf{E}_{\omega }$ and
$\mathbf{B}_{\omega }$ with
respect to $\mathbf{E}_{0}$ and $\mathbf{B}_{0}$\ changes in the $%
K^{^{\prime }}$ frame. We know that

\begin{equation}
E^{2}-B^{2}=E^{\prime 2}-B^{\prime 2}  \label{TransTotal}
\end{equation}

\begin{equation}
\mathbf{{E}^{\prime }\cdot {B}^{\prime }={E}\cdot {B}}
\label{ScalarTotal}
\end{equation}%
We also have that $E_{0}^{2}-B_{0}^{2}=E_{0}^{^{\prime
}2}-B_{0}^{\prime 2}$
and $\mathbf{E}_{0}^{\prime }\cdot \mathbf{B}_{0}^{\prime }\mathbf{=E}%
_{0}\cdot \mathbf{B}_{0}$ (actually in our specific case
$B_{0}^{^{\prime }}=0$ (resp. $E_{0}^{^{\prime }}=0$)). We finally
derive from Eq. \ref{TransTotal}
that%
\begin{equation}
\mathbf{{E}_{\omega }\cdot {E}_{0}-{B}_{\omega }\cdot
{B}_{0}={E}_{\omega
}^{\prime }\cdot {E}_{0}^{\prime }\;(\mbox{resp. } {B}_{\omega }^{\prime }\cdot {B}%
_{0}^{\prime })}  \label{Rel1}
\end{equation}%
and, from Eq. \ref{ScalarTotal} that%
\begin{equation}
\mathbf{{E}_{\omega }\cdot {B}_{0}+{E}_{0}\cdot {B}_{\omega }={E}%
_{0}^{\prime }\cdot {B}_{\omega }^{\prime }\;(\mbox{resp. }
{E}_{\omega }^{\prime }\cdot {B}_{0}^{\prime })}  \label{Rel2}
\end{equation}%
If for example $\mathbf{E}_{\omega }\parallel \mathbf{E}_{0}$ and $\mathbf{B}%
_{\omega }\parallel \mathbf{B}_{0}$\ Eq. \ref{Rel2} gives that $\mathbf{E}%
_{0}^{\prime }\cdot \mathbf{B}_{\omega }^{\prime }=0$ i.e. $\mathbf{E}%
_{\omega }^{\prime }\parallel \mathbf{E}_{0}^{\prime }$. If vice versa $%
\mathbf{E}_{\omega }\parallel \mathbf{B}_{0}$ and
$\mathbf{B}_{\omega }\parallel \mathbf{E}_{0}$ Eq. \ref{Rel1}
gives that $\mathbf{E}_{0}^{\prime }\cdot \mathbf{E}_{\omega
}^{\prime }=0$ i.e. $\mathbf{B}_{\omega }^{\prime }\parallel
\mathbf{E}_{0}^{\prime }$. Summarizing, the orientation of the
polarization of the wave with respect to the only existing static
field in the $K^{^{\prime }}$ frame is the same with respect to
that static field in the $K$ frame.

\subsection*{Results}

Let's finally calculate the value of $n$ using the expressions
derived in the previous paragraphs. We consider the situation when
$\mathbf{E}_{0}\bot
\mathbf{B}_{0}$ and assume for the moment that $E_{0}<B_{0}$ and that $%
\mathbf{k}$ is parallel to $\mathbf{E}_{0}\times \mathbf{B}_{0}$
and therefore parallel to $\mathbf{v}$. In the frame $K^{^{\prime
}}$, the only existing effect is a Cotton-Mouton effect
proportional to the square of $B^{\prime}_0$ i.e. $\delta
n^{\prime }$ depends on the polarization of
light. Thanks to eqs. $\ref{npar}$, $\ref{nperp}$, we can write that%
\begin{equation}
\delta n_{\Vert }^{\prime }=\eta_{\Vert }B_{0}^{\prime
2}=\eta_{\Vert }(1-\beta ^{2})B_{0}^{2}
\end{equation}%
and, using Eq. \ref{Fizeau1},%
\begin{equation}
\delta n_{\Vert }=\eta_{\Vert }(1-\beta )^{2}B_{0}^{2}=\eta_{\Vert
}(B_{0}^{2}-2E_{0}B_{0}+E_{0}^{2})
\end{equation}%
since $\mathbf{E}_{0}$ and $\mathbf{B}_{0}$ are perpendicular. For
the same reason, we can write
\begin{equation}
\delta n_{\bot }=\eta_{\bot}(1-\beta
)^{2}B_{0}^{2}=\eta_{\bot}(B_{0}^{2}-2E_{0}B_{0}+E_{0}^{2})
\end{equation}%
So if $\eta_{\Vert }\neq \eta_{\bot}$ i.e. if the Cotton-Mouton
effect exists, we have demonstrated that the Kerr effect and the
magneto-electric birefringence proportional to ${E}_{0}{B}_{0}$
must also exist. Moreover, because of Lorentz invariance, Kerr
birefringence has to be of opposite sign compared to the
Cotton-Mouton one since light retardation is equal for light
polarized parallel to the $\mathbf{B}_{0}$ field and orthogonal to the $%
\mathbf{E}_{0}$ field and vice versa. As for magneto-electric
birefringence, the coefficient that multiplies the fields is twice
the one of the Cotton-Mouton or Kerr birefringence.

Let's now come to the polarization-independent magneto-electric
anisotropy. If we write $\eta_{\Vert }=\eta_{\bot}+\Delta $ we
find that light is retarded by the existence of the static fields
independently of the polarization by a quantity corresponding to
\begin{equation}
\delta n_{0}(k)=\eta_{\bot}(B_{0}^{2}-2E_{0}B_{0}+E_{0}^{2})
\end{equation}%
If we change $\mathbf{k}$ into $-\mathbf{k}$, we have to change the sign of $%
\beta $ and therefore the previous equation becomes
\begin{equation}
\delta n_{0}(-k)=\eta_{\bot}(B_{0}^{2}+2E_{0}B_{0}+E_{0}^{2})
\end{equation}%
and finally we find for the anisotropy
\begin{equation}
\delta n_{0}(k)-\delta n_{0}(-k)=-4\eta_{\bot}E_{0}B_{0}
\end{equation}%
For the case when $B_{0}<E_{0}$, one obtains exactly the same
results and
they are identical to those that we have obtained in ref. \cite%
{RikkenandRizzo} and that we have complemented in ref. \cite%
{RikkenandRizzoBR} where we have used for $\eta_{\Vert }$ and
$\eta_{\bot}$ the accepted values. We stress that the existence of
the magneto-electric anisotropy is only related to the fact that
$\eta_{\bot}$ is different from zero. This effect could exist even
if the Cotton-Mouton effect would not exist, as the Cotton-Mouton
effect exists only if $\Delta $ is different from zero,
independent of the value of $\eta_{\bot}$.

We finally study the case when $E_{0} = B_{0}$. If $\mathbf{k}$ is
parallel to ${\mathbf{E}_{0}\times \mathbf{B}_{0}}$, $\delta
n_{\Vert }=\delta n_{\bot }=0$. This result is somewhat obvious
since this case corresponds to the propagation of a plane wave in
the field of a co-propagating plane wave. If $\mathbf{k}$ is
antiparallel to ${\mathbf{E}_{0}\times \mathbf{B}_{0}}$, the
effect is not zero. This means that two counterpropagating plane
waves can affect each other.

We note that $\eta_{\Vert }$ and $\eta_{\bot}$ are directly
related to the coefficients of the invariants $F^{2}$ and $G^{2}$
in the Lagrangian $L_{EK}$ \cite{Zavattini}. Actually,
$\eta_{\Vert }={\frac{b}{2}}$ and $\eta_{\bot}=2a$ and, for the
Cotton Mouton effect, we have

\begin{equation}
\Delta n=n_{\Vert }-n_{\bot }={\frac{1}{2}}(b-4a)B_{0}^{2}
\end{equation}%
i.e. $\Delta n={\frac{3}{2}}aB_{0}^{2}$ since $b=7a$. The
existence of the Cotton-Mouton effect in vacuum therefore depends
on the ratio ${b/a}$. If for example ${b/a}$ would be equal to
$4$, $\Delta n$ would be zero and no Cotton-Mouton effect would
exist. Nor would the Kerr effect, the magneto-electric
birefringence or the magneto-electric Jones birefringence exist.
The only existing effect would be the polarization-independent
magneto-electric anisotropy. This fact is not without importance
at least from a historical point of view, since Born and Infeld
have developed around 1934 a QED theory \cite{BornInfeld} in which
the value predicted for the ratio ${b/a}$ was exactly $4$.

The case of $\mathbf{E}_{0}$ perpendicular to $\mathbf{B}_{0}$ with $\mathbf{%
E}_{0}$ or $\mathbf{B}_{0}$ parallel to $\mathbf{k}$ has not been
treated before and we will show in the following how we can solve
it using our method. Let's assume that $E_{0}<B_{0}$ and consider
$\mathbf{B}_{0}\Vert \mathbf{k}$. The velocity $\mathbf{v}$ is
perpendicular to $\mathbf{k}$ in frame $K$. In the frame
$K^{^{\prime }}$, the $\mathbf{B}_{0}^{\prime }$ field lies in the plane containing $\mathbf{k}^{\prime }$ and $%
\mathbf{v}$ and it is perpendicular to $\mathbf{v}$. The
$\mathbf{k}^{\prime }$
vector is no more perpendicular to $\mathbf{v}$ and therefore $\mathbf{B}%
_{0}^{\prime }$ is no more collinear with the direction of
propagation of light. In this frame, only the Cotton-Mouton effect
can exist, and to calculate $\delta n{^{\prime }}$, one has
therefore only to consider the component of
$\mathbf{B}_{0}^{\prime }$ perpendicular to $\mathbf{k}^{\prime
}$. Using the result for $\cos(\theta ^{^{\prime }}),$ this
component ${B}_{0,\perp }^{\prime }$ is equal to

\begin{equation}
{B}_{0,\perp }^{\prime }=-\beta {B}_{0}^{\prime }
\end{equation}%
and%
\begin{equation}
\delta n^{^{\prime }}\propto (1-\beta ^{2})B_{0}^{2}\beta ^{2}
\end{equation}%
and finally, using Eq. \ref{Fizeau3},%
\begin{equation}
\delta n\propto B_{0}^{2}\beta ^{2}=E_{0}^{2}
\end{equation}%
Therefore only the Kerr effect exists for this case. Let's now
consider the case $E_{0}>B_{0}$. The resulting
$\mathbf{E}_{0}^{\prime }$ is still perpendicular to $\mathbf{v}$
and $\mathbf{k}^{\prime }$. This means that in $K^{^{\prime }}$,
the only existing effect is a Kerr effect given by the entire
electric field.

\begin{equation}
\delta n^{^{\prime }}\propto (1-\beta ^{2})E_{0}^{2}
\end{equation}%
and finally by Eq. \ref{Fizeau3},

\begin{equation}
\delta n\propto E_{0}^{2}
\end{equation}
Again, the $\mathbf{B}_{0}$ field gives no contribution. It is
straightforward to show that in order to solve the case when the $\mathbf{E}%
_{0}$ field is parallel to $\mathbf{k}$, it is sufficient to permute $%
\mathbf{E}_{0}$ with $\mathbf{B}_{0}$ in the previous formulas.
Therefore, in general, the static field parallel to $\mathbf{k}$
does not contribute to a bilinear optical effect.

\subsection*{Parallel fields geometry}

To complete our analysis, we consider the geometry where the two fields $%
\mathbf{B}_{0}$ and $\mathbf{E}_{0}$ are parallel to each other.
The case in which the two fields are perpendicular to the
$\mathbf{k}$ vector of light
has been studied in detail by Ross, Sherborne and Stedman in ref. \cite%
{Stedman}. These authors proved by symmetry considerations that in
this configuration, magneto-electric Jones birefringence should
exist and that it should have the same magnitude as the
magneto-electric linear birefringence in crossed fields. The two
effects are actually two different facets of the same phenomenon.

What has not been studied yet is the case where the two parallel fields are parallel to the $%
\mathbf{k}$ vector. We will show that no bilinear effect exists
for this
geometry. First assume that such an effect exists i.e.%
\[
\delta n=\eta E_{0}B_{0}
\]%
Let's regard the two fields in a $K^{\prime }$ reference frame
moving at a velocity $\mbox{\boldmath $\beta$ }$ in the direction
of $\mathbf{k}$. Using equations
\ref{LorTransE'} and \ref{LorTransB'}, one can show that%
\begin{equation}
\mathbf{{E}_{0}^{\prime }={E}_{0}}
\end{equation}%
and%
\begin{equation}
\mathbf{{B}_{0}^{\prime }={B}_{0}}
\end{equation}%
In $K^{\prime }$ therefore%
\[
\delta n^{\prime }=\eta {E}_{0}{B}_{0}
\]%
Using equation $\ref{Fizeau1}$ we must then conclude that%
\begin{equation}
\delta n=\eta E_{0}B_{0}{\frac{{1-\beta }}{{1+\beta }}}
\end{equation}%
Since $\beta $ can take any value between 0 and 1,this is in
disagreement with our starting assumption, unless $\eta$ is equal
to zero. Thus the two parallel fields in the direction of
$\mathbf{k}$ give no bilinear effect.

\subsection*{Conclusion}

We have shown that the magneto--electric optical properties of the
quantum vacuum can be deduced from the principle of Lorentz
invariance. We have reproduced in a straightforward manner
theoretical results obtained by much more elaborate methods, and
have also considered new geometries, never treated before. In
particular we have found the new result that in crossed static
electric and magnetic fields, a static component parallel to the
propagation of light does not give rise to an optical effect.
Although our method could also be applied to material media, it is
less convenient there, as one also has to take into account the
transformations of the induced material responses.

\subsection*{Acknowledgements}

The authors are grateful to C. Robilliard for her comments and
careful reading of the manuscript.

\end{document}